\documentclass[epjCONF,columns]{svjour} % for 2 columns format
\usepackage{graphics}
\usepackage{graphicx}
\usepackage[varg]{txfonts} % Times fonts
\usepackage[latin1]{inputenc}
\usepackage{lineno}
\usepackage{subfig}
\usepackage{xspace}
\usepackage{floatrow}
\newcommand{\ttbar}{\ensuremath{t\overline{t}}\xspace}
\newcommand{\MC}{Monte~Carlo\xspace}
\newcommand{\MET}{\ensuremath{E\!\!\!\slash_t}\xspace}
\newcommand{\MTW}{\ensuremath{M_T(W)}\xspace}
\newcommand{\qb}{\ensuremath{b}\xspace}
\newcommand{\W}{\ensuremath{W}\xspace}
\newcommand{\Wjets}{\ensuremath{W\!+}jets\xspace}
\newcommand{\Wc}{\ensuremath{W\!\!+\!\!c+}jets\xspace}
\newcommand{\Z}{\ensuremath{Z}\xspace}
\session-title{2011 Hadron Collider Physics symposium (HCP-2011)}
\begin{document}
%
%\linenumbers % REMOVE BEFORE SUBMISSION TO JOURNAL!!!
%
\title{Single Top Results by ATLAS and CMS}
\author{Dr. Petra Haefner\thanks{\email{Petra.Haefner@cern.ch}}, On behalf of the ATLAS and CMS Collaborations}
\institute{Max-Planck-Institut f\"ur Physik, M\"unchen}
\abstract{
Electroweak production of top quarks, the so-called single top production, is interesting both in the context of measurements of the Standard Model of Particle Physics as well as searches for new phenomena beyond that. Analyses based on data taken in 2010 and 2011 by the ATLAS and CMS collaborations at a centre-of-mass energy of 7~TeV at the LHC proton-proton collider will be presented. Different production channels, including t-channel, s-channel, and associated \W boson production, have been addressed using datasets of up to 2.1~fb$^{-1}$. A first search for Flavour Changing Neutral Currents has been performed as well.
} %end of abstract
\maketitle

%%%%%%%%%%%%%%%%%%%%%%%%%%%%%%%%%%%%%%%%%%%%

\section{Introduction}
\label{intro}
\begin{figure}
\centering
\resizebox{\columnwidth}{!}{
\subfloat[]{
   \includegraphics[width=0.2\columnwidth]{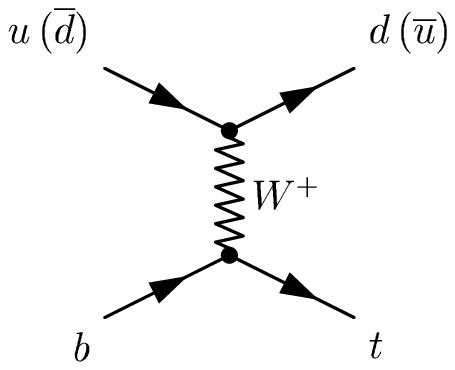}
   \label{fig:feyn_t}
 }
 \subfloat[]{
   \includegraphics[width=0.15\columnwidth]{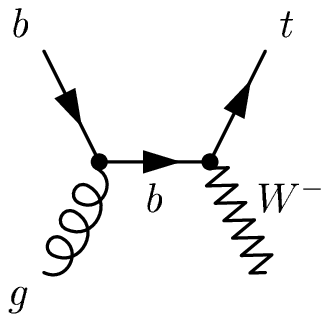}
   \label{fig:feyn_Wt}
 }
 \subfloat[]{
   \includegraphics[width=0.15\columnwidth]{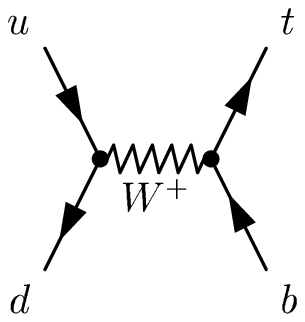}
   \label{fig:feyn_s}
 }
 \subfloat[]{
   \includegraphics[width=0.3\columnwidth]{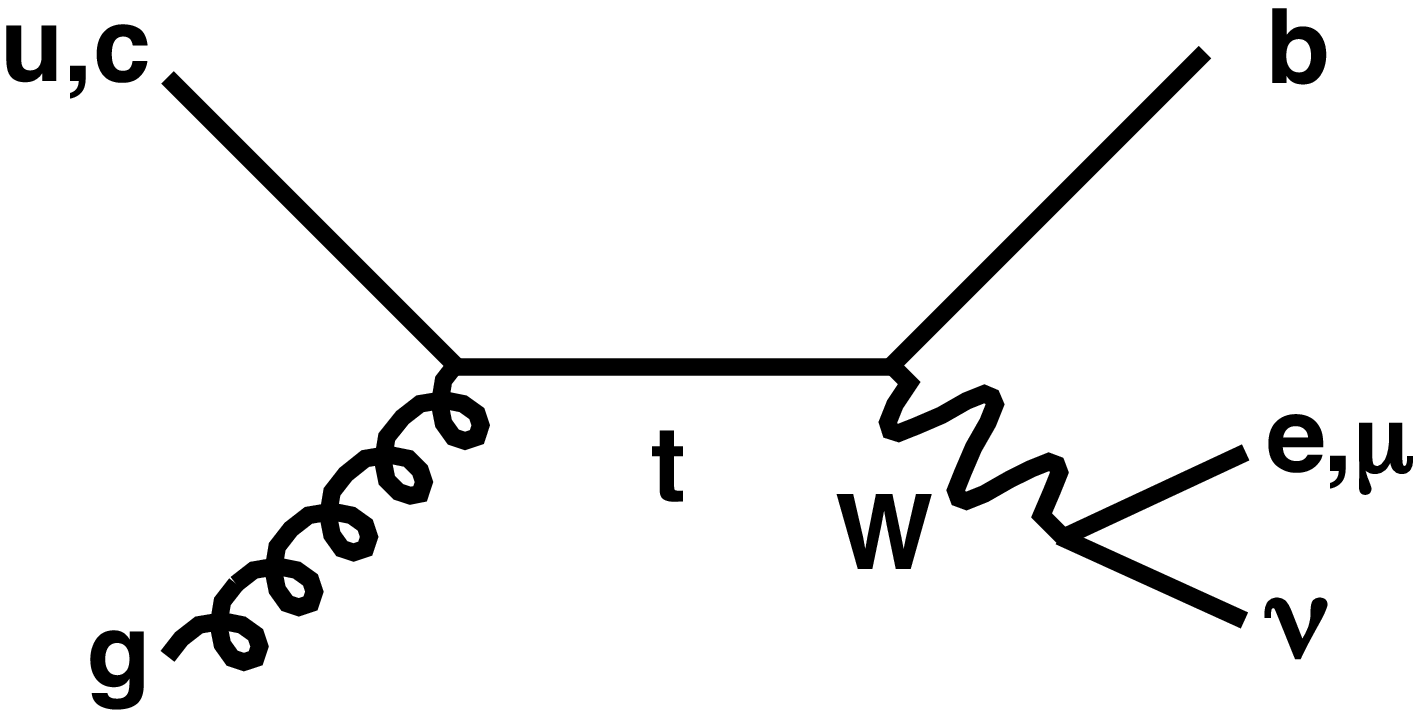}
   \label{fig:feyn_FCNC}
 }
 }
\vspace{-0.25cm}
\caption{Examples of Feynman diagrams for \protect\subref{fig:feyn_t} t-channel, \protect\subref{fig:feyn_Wt} Wt-channel, \protect\subref{fig:feyn_s} s-channel single top quark production. \protect\subref{fig:feyn_FCNC} FCNC single top quark production $qg \rightarrow t$ with a Standard Model top decay $t \rightarrow Wb \rightarrow l\nu\, b$.}
\label{fig:feynman}
\end{figure}

In the Standard Model (SM) of Particle Physics, top quarks can be produced either in top-antitop pairs via the strong interaction or as single top quarks by the electroweak interaction. Top quarks decay (in the SM) almost exclusively to a \W boson and a \qb quark. Therefore, the decay signature depends solely on the \W, decaying  either leptonically into lepton+neutrino $l\nu$ or hadronically into quark-antiquark $q\overline q$. In general, cross sections are lower and backgrounds higher for single top quark production compared to top pairs making analyses more difficult. Therefore, single top production was established only in recent years at the Tevatron. At the higher centre-of-mass energy at the LHC, cross sections are larger for the different production channels, leading to higher statistics. Figure~\ref{fig:feynman} shows examples of Feynman diagrams for t-channel, associated \W boson production (Wt-channel) and s-channel single top production. Table~\ref{tab:xsec} lists the cross sections at Next-to-leading order (NLO) with Next-to-next-to- leading-log (NNLL) resummation (NNLO approximation) for Tevatron and LHC centre-of-mass energies \cite{xsec_1,xsec_2,xsec_3}. Cross sections at the LHC are between about 4 and 30 times larger than at the Tevatron. Especially the Wt-channel, which was not accessible before, has a significant cross section at the LHC. Even the s-channel cross section, the smallest one at the LHC, is more than two times larger than the largest one at the Tevatron. This allows detailed measurements of the Standard Model and searches for new phenomena in the coming years.
The different production channels will be established, their cross sections compared to Standard Model predictions, the unitarity of the CKM matrix can be tested and b-quark Parton Density Functions studied. Possible searches include signatures of Flavour Changing Neutral Currents (FCNCs) (cf. Fig.~\ref{fig:feyn_FCNC}), \W' boson, charged Higgs H$^+$, a fourth generation of quarks and so on. Part of that program is already possible with the current ATLAS and CMS datasets of up to 2~fb$^{-1}$. More signatures will be studied in the future.

\begin{table}
\begin{tabular}{c|r@{$\pm$}lr@{$\pm$}lr@{$\pm$}l}
\hline
\noalign{\smallskip}
$\sqrt{s}$ & \multicolumn{2}{c}{t-channel} & \multicolumn{2}{c}{Wt-channel} & \multicolumn{2}{c}{s-channel} \\
\noalign{\smallskip}
\hline
\noalign{\smallskip}
7~TeV & 64.2 & 2.6~pb & 15.6 & 1.3~pb & 4.6 & 0.2~pb\\
1.96~TeV & 2.1 & 0.1~pb & 0.25 & 0.03~pb & 1.05 & 0.05~pb\\
\noalign{\smallskip}
\hline
\end{tabular}
\caption{Cross sections at NLO+NNLL for single top production at LHC and Tevatron centre-of-mass energies \protect\cite{xsec_1,xsec_2,xsec_3}. A top quark mass of 173~GeV is assumed.}
\label{tab:xsec} 
\end{table}

%%%%%%%%%%%%%%%%%%%%%%%%%%%%%%%%%%%%%%%%%%%%

\section{Measurements in the t-channel}
\label{sec:tchannel}

Due to its large cross section, the t-channel single top quark production has been addressed first by the ATLAS and CMS collaborations. Both have presented measurements based on the first year of LHC data, corresponding to 36~pb$^{-1}$ (CMS) \cite{CMS_t}  and 35~pb$^{-1}$ (ATLAS)  \cite{ATL_t_1}. The ATLAS collaboration has updated this analysis with 2011 data corresponding to 156~pb$^{-1}$ \cite{ATL_t_2} and a second time with 700~pb$^{-1}$ \cite{ATL_t_3}. Only the latter analysis will be described in the following. Both collaborations have used cut-based measurement techniques which are simple and robust as well as multivariate approaches which have in general higher sensitivities as they exploit the full kinematics of signal and backgrounds. However, those techniques have larger model dependencies. CMS uses a Boosted Decision Tree (BDT) approach, whereas ATLAS applies a Neural Network (NN) technique. 

Both collaborations only consider events, where the \W boson originating in the top quark decays leptonically into a lepton and neutrino. Therefore, an isolated lepton, either an electron or a muon, is required in the event selection. They use anti-k$_T$ jets with a size parameter $\Delta R$ of 0.5 for CMS and 0.4 for the ATLAS analysis. CMS selects only events with exactly 2 jets whereas ATLAS also allows 3-jet events. Both analysis teams require one jet to be identified as a \qb-jet. CMS uses a technique based on the interaction point significance, whereas ATLAS explicitly identifies a secondary vertex, indicating a \qb-decay. Both collaborations also make requirements on the \W boson from the top decay. CMS asks for a transverse mass of the lepton and the neutrino to be larger than 40~GeV in the muon case and larger than 50~GeV in the electron case. ATLAS requires that the transverse lepton-neutrino mass is larger than 60~GeV minus the missing transverse energy (\MET). In addition, ATLAS makes an explicit cut on \MET to be larger than 25~GeV. More details about the event selection can be found in \cite{CMS_t,ATL_t_3}.

\begin{figure}
	\begin{center}
		\resizebox{0.49\columnwidth}{!}{
			\includegraphics{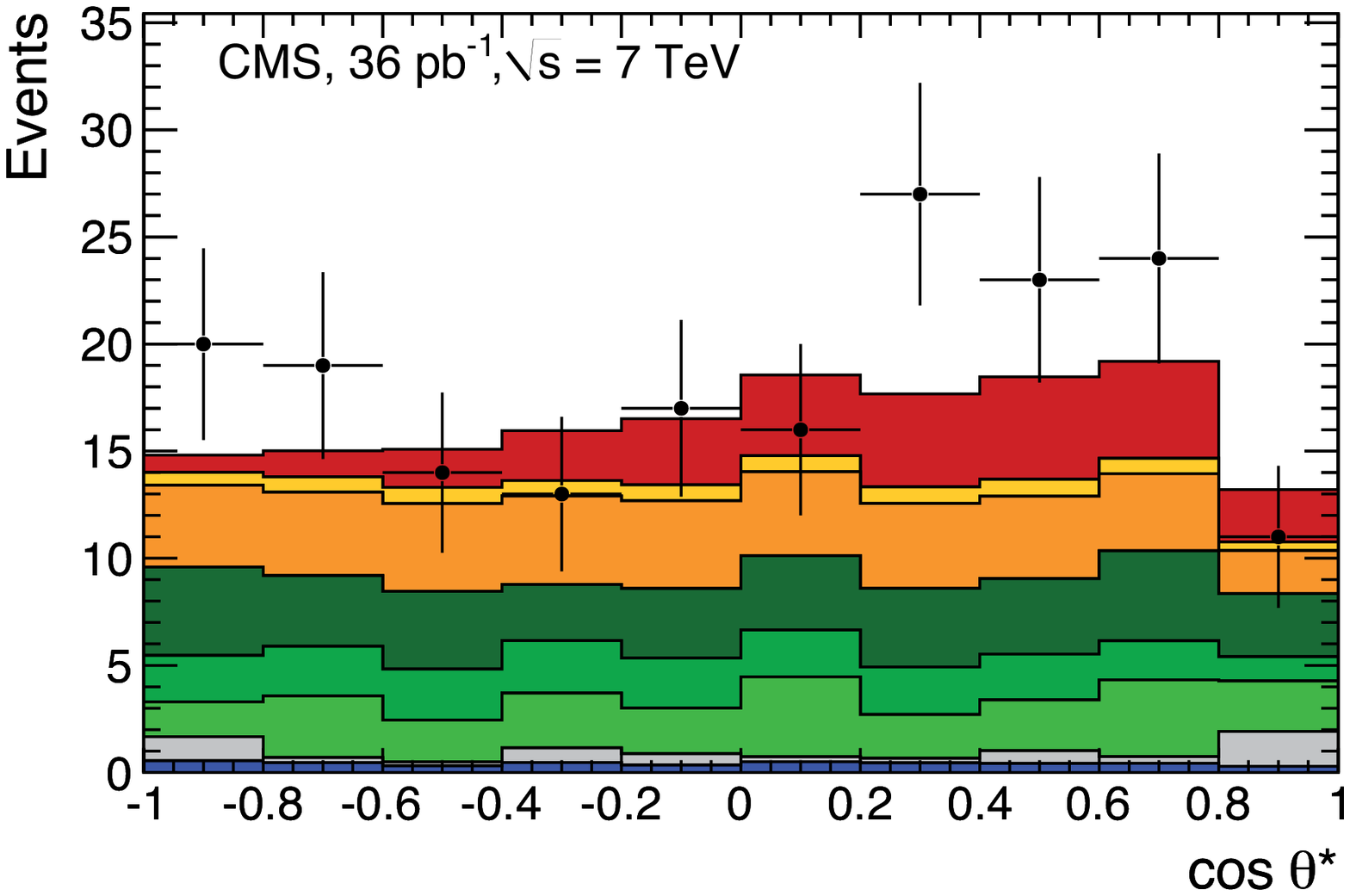}}
		\resizebox{0.49\columnwidth}{!}{
			\includegraphics{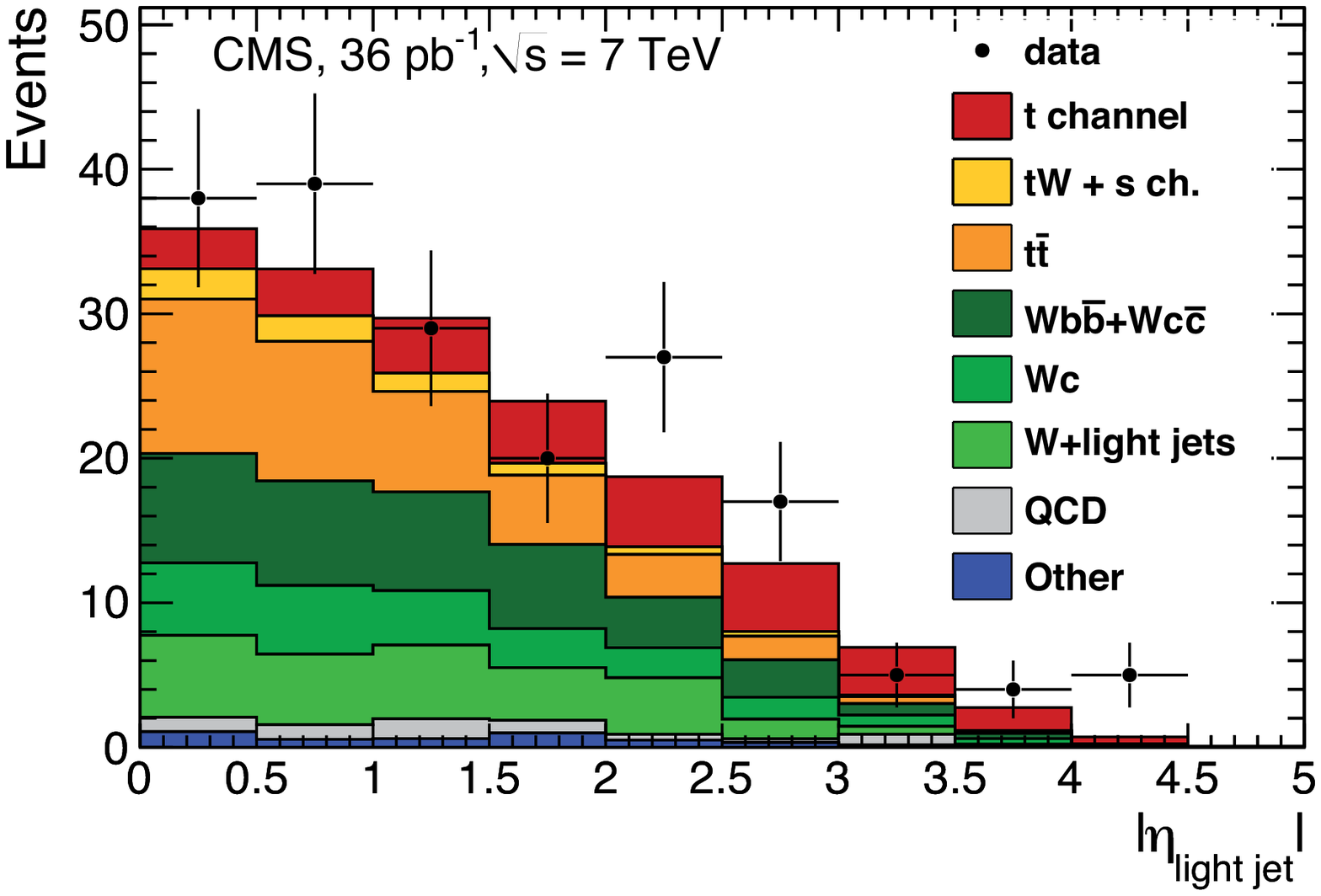}}
		\vspace{-0.5cm}
		\caption{Result of the CMS t-channel cut-based analysis \protect\cite{CMS_t}. \emph{Left:} cos-distribution of the angle between the lepton and light jet. \emph{Right:} Distribution of the absolute pseudorapidity of the light jet. A 2-dim. fit of both distributions with signal and background as free parameters is performed.}
		\label{fig:CMS_t_cut} 
	\end{center}
\end{figure}
\begin{figure}
	\begin{center}
			\resizebox{0.49\columnwidth}{!}{
				\includegraphics{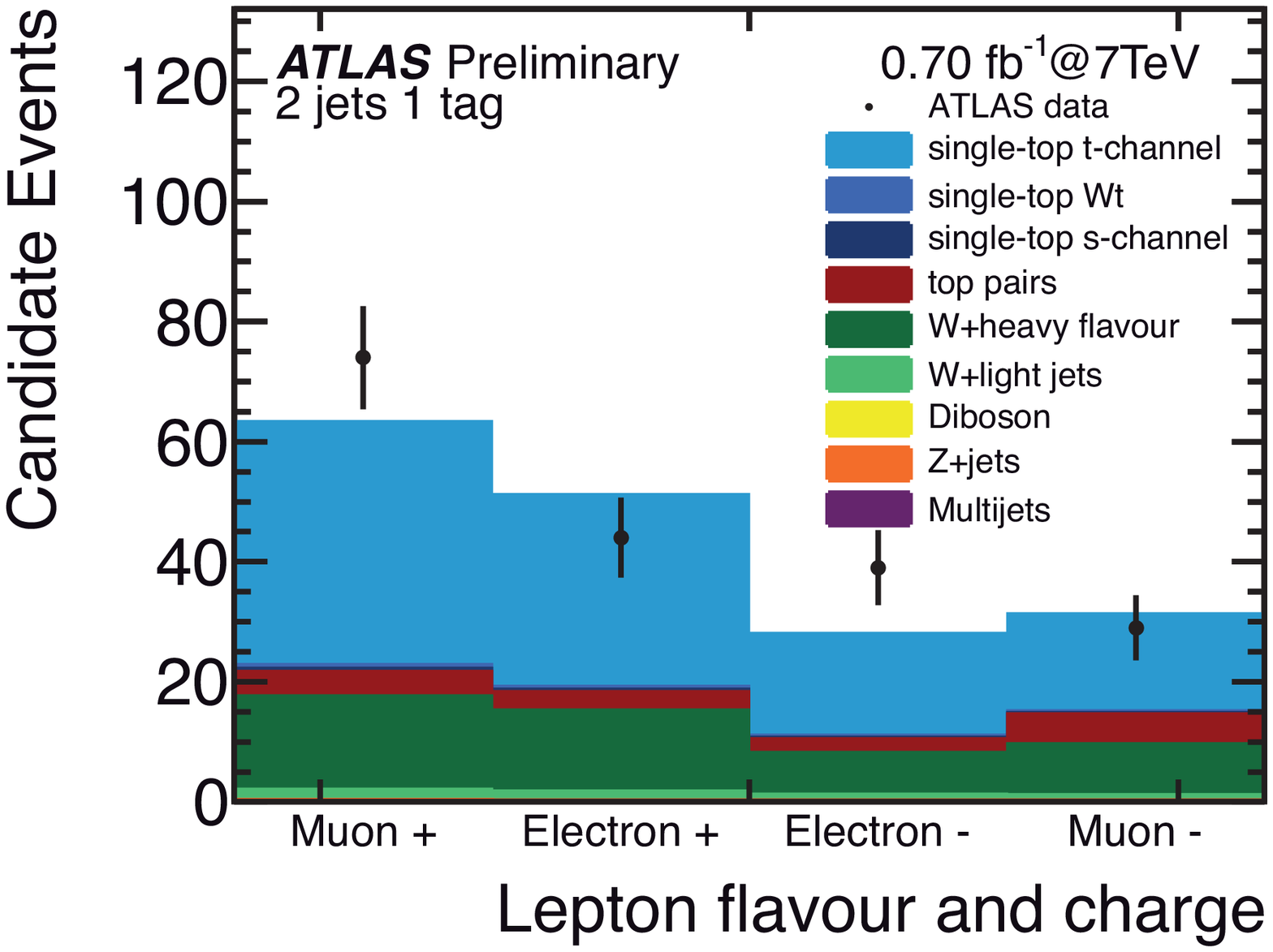}}
			\resizebox{0.49\columnwidth}{!}{	
				\includegraphics{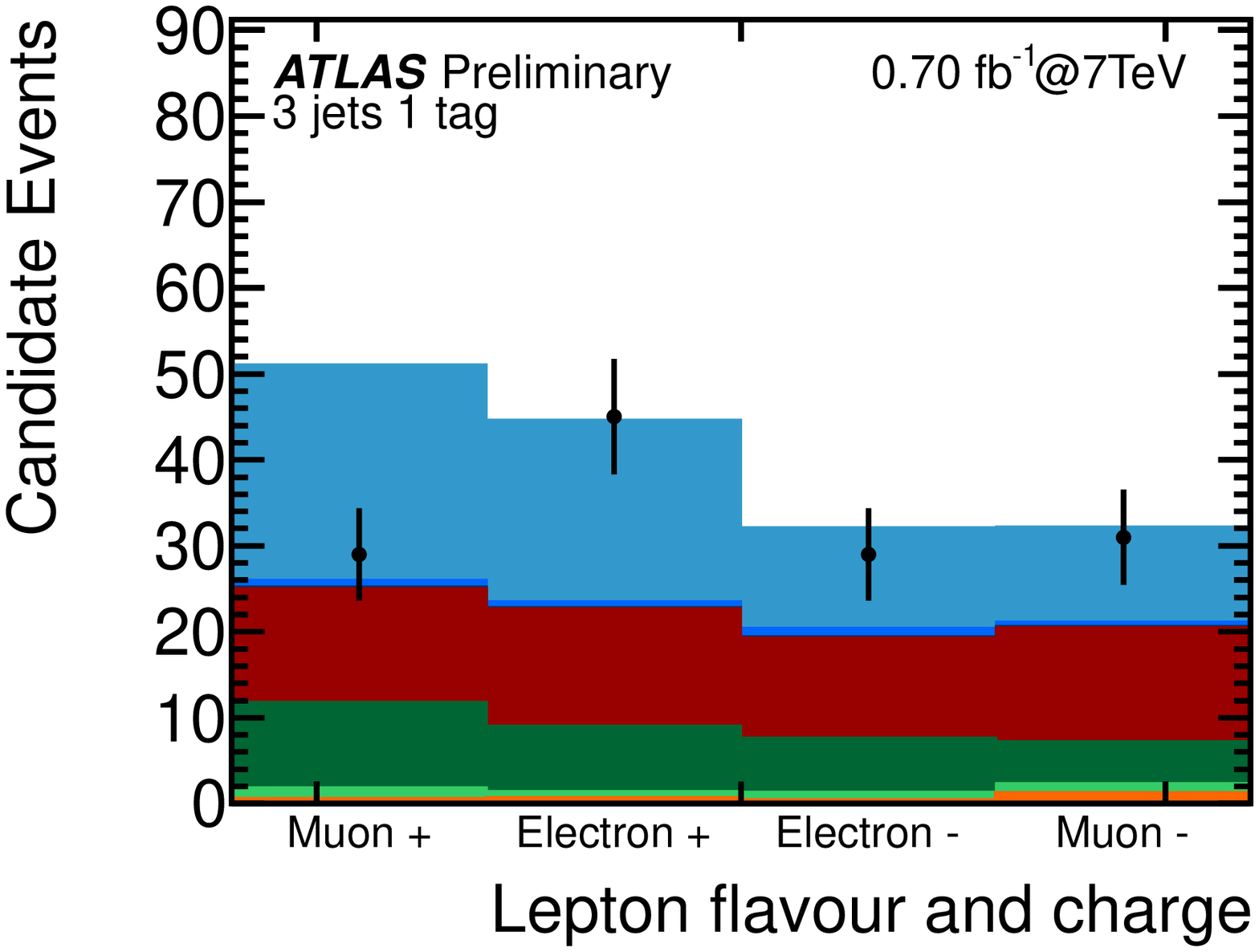}}
		\vspace{-0.5cm}
		\caption{Result of the fit to 8 subsamples in the 2- and 3-jet signature of the ATLAS t-channel cut-based analysis \protect\cite{ATL_t_3}. Signal and background are allowed to vary freely in the fit.}
		\label{fig:ATL_t_cut} 
	\end{center}
\end{figure}
\begin{figure}
	\begin{center}
		\resizebox{0.49\columnwidth}{!}{
			\includegraphics{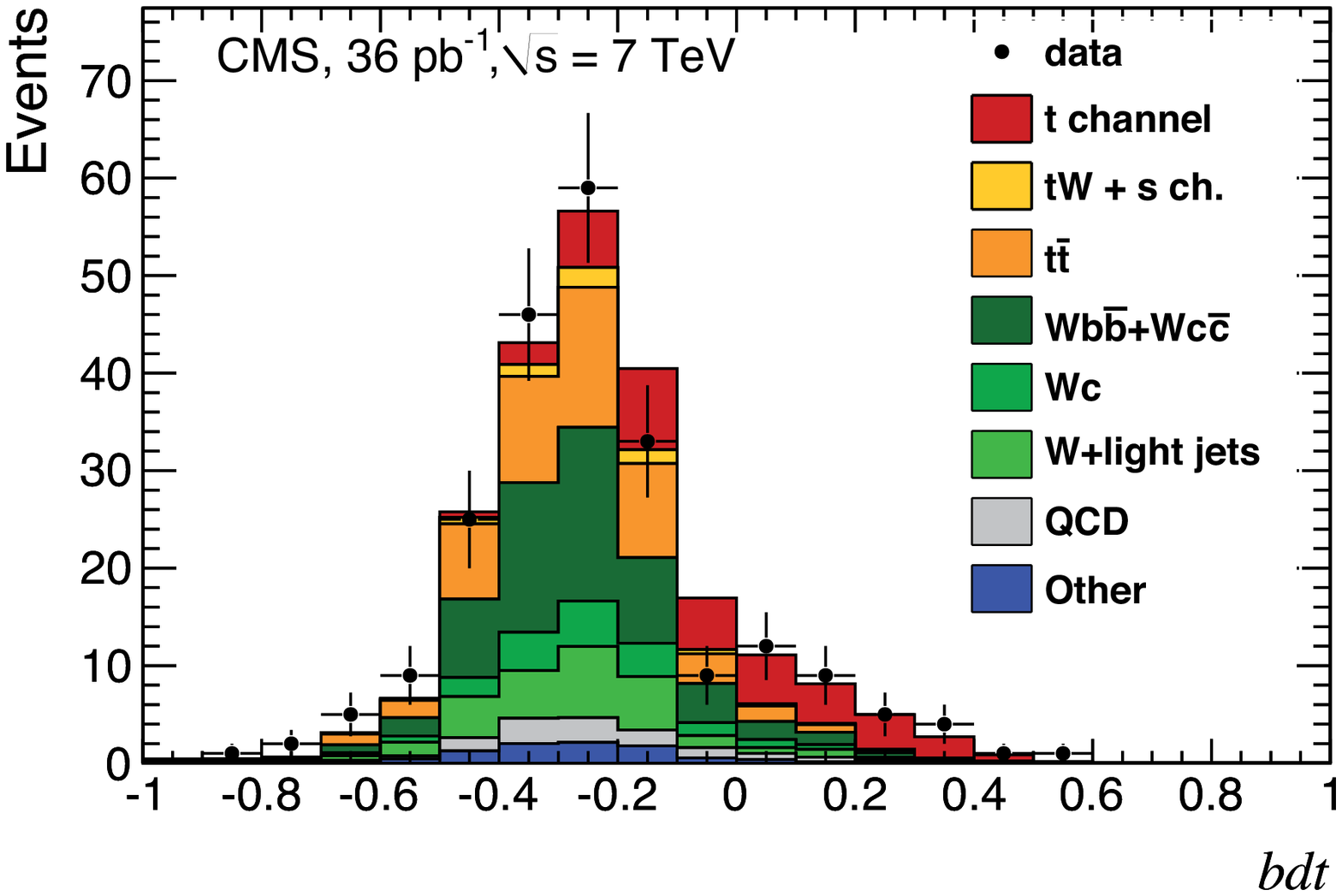}}
		\resizebox{0.49\columnwidth}{!}{	
			\includegraphics{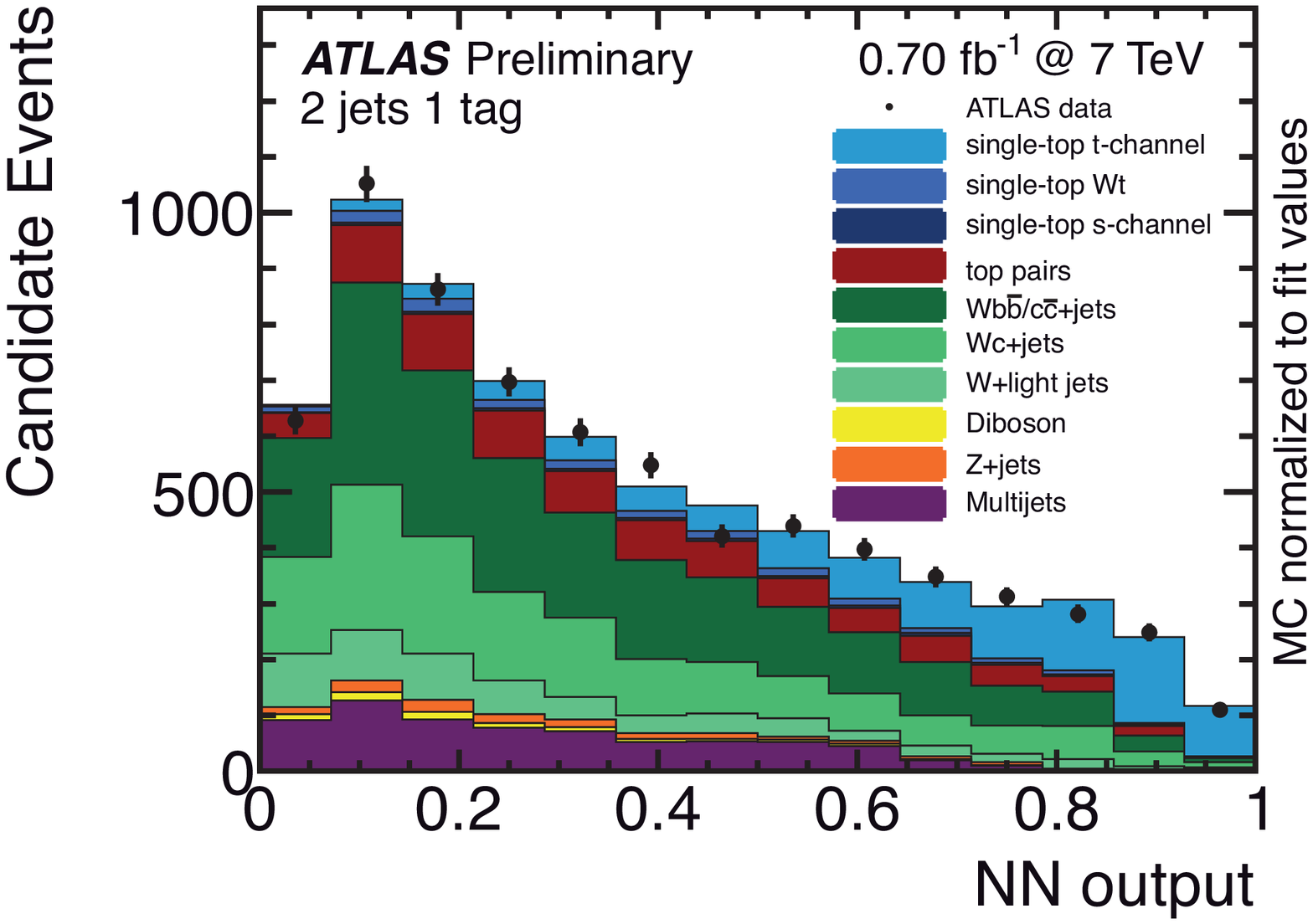}}
		\vspace{-0.5cm}
		\caption{Results of t-channel multivariate analyses. \emph{Left:} CMS BDT output \protect\cite{CMS_t}. \emph{Right:} ATLAS NN output \protect\cite{ATL_t_3}.}
		\label{fig:t_MV} 
	\end{center}
\end{figure}

After this event selection, the most important backgrounds are \ttbar pair production, \W bosons in association with jets (\Wjets) and QCD multijet events. ATLAS normalizes the \ttbar background to the theory prediction, CMS to the measured value of 150~pb. For the \Wjets background ATLAS takes the shape from \MC simulation. The size is estimated from a simultaneous fit to the NN output or in the cut-based analysis taken from scale factors derived in three control regions. CMS distinguishes \Wjets background containing heavy flavour jets (HF) or light jets only. For the first LO predictions are used scaled by factors of the \ttbar cross section analysis. The latter is scaled to NNLO precision in the BDT analysis. The cut-based analysis uses a normalization fitted to the transverse mass of the W boson \MTW in two control regions. The shape is estimated by a data-driven model. For the QCD multi jet background ATLAS performs a likelihood fit to the \MET distribution to estimate the size and takes the shapes from a jet-electron model and a loose isolation muon sample. CMS measures the size in a likelihood fit to the \MTW distribution and the shape from orthogonal lepton isolation samples.

In the cut-based approach, CMS performs a maximum likelihood fit in two variables, which are the cosine of the angle between the light jet (i.e. the non-\qb-identified jet) and the lepton $cos~\theta^\ast$ and the absolute value of the pseudorapidity of the light jet $|\eta_{light jet}|$. These variables are chosen in order to minimize model dependencies. The background is treated as a free parameter in the fit. The data distributions of the two variables together with the results of the fit are shown in Fig.~\ref{fig:CMS_t_cut}. As stated before, ATLAS uses both the 2-jets (1 \qb-tag) and 3-jets (1 \qb-tag) samples. Both samples are separated into four subsamples of lepton flavour and charge. These eight subsamples are simultaneously fitted. As in the CMS analysis, the background is a free parameter. The result of the fit together with the data points for the eight subsamples is given in Fig.~\ref{fig:ATL_t_cut}

In the multivariate analyses, a Boosted Decision Tree (BDT) and a Neural Network (NN) are used by the CMS and ATLAS collaborations, respectively. Both analysis teams perform a fit to the full output distribution and again include the background as free parameter. In the ATLAS analysis it should be noted that only the 2-jet sample is used, therefore the statistics are lower than in the respective cut-based analysis. Results of the fits are shown in Fig.~\ref{fig:t_MV}.

CMS measures a cross section of $124\pm34^{+30}_{-34}$~pb in the cut-based analysis and $79\pm25^{+13}_{-15}$~pb in the BDT analysis. The first uncertainty is statistical whereas the second one is systematic. The dominant systematic uncertainty is the \qb-identification efficiency. Other major systematics are the signal model, the factorization and renormalization scales of the \Wjets background, the jet energy scale and the size of the \Wc background. A combination of the two measurements yields a cross section of $84\pm30$~pb for the t-channel single top production. In a background only hypothesis, the deviation between measurement and expected background corresponds to 3.5\,$\sigma$. CMS derives a lower limit for the CKM matrix element $|V_{tb}|$ of 0.62 in the cut-based and 0.68 in the BDT approach. The estimate is restricted to a range of $0 \le |V_{tb}| \le 1$. %??? Clarify!!!

ATLAS measures the cross section of t-channel single top production to be $90\pm9^{+31}_{-20}$~pb in the cut-based approach, corresponding to 7.6\,$\sigma$. The NN analysis on the smaller dataset of 2-jet events only serves as a cross check and yields $105\pm7^{+36}_{-30}$~pb. As for CMS \qb-identification is the dominant systematic uncertainty, expressed in the \qb-tagging scale factor. Other major systematics are the jet pseudorapidity reweighing, the choice of \MC generator, the parton shower model and the jet energy scale.

%%%%%%%%%%%%%%%%%%%%%%%%%%%%%%%%%%%%%%%%%%%%
\section{Measurements in the Wt-channel}
\label{sec:Wtchannel}

\begin{figure}
	\begin{center}
			\resizebox{0.45\columnwidth}{!}{
				\includegraphics{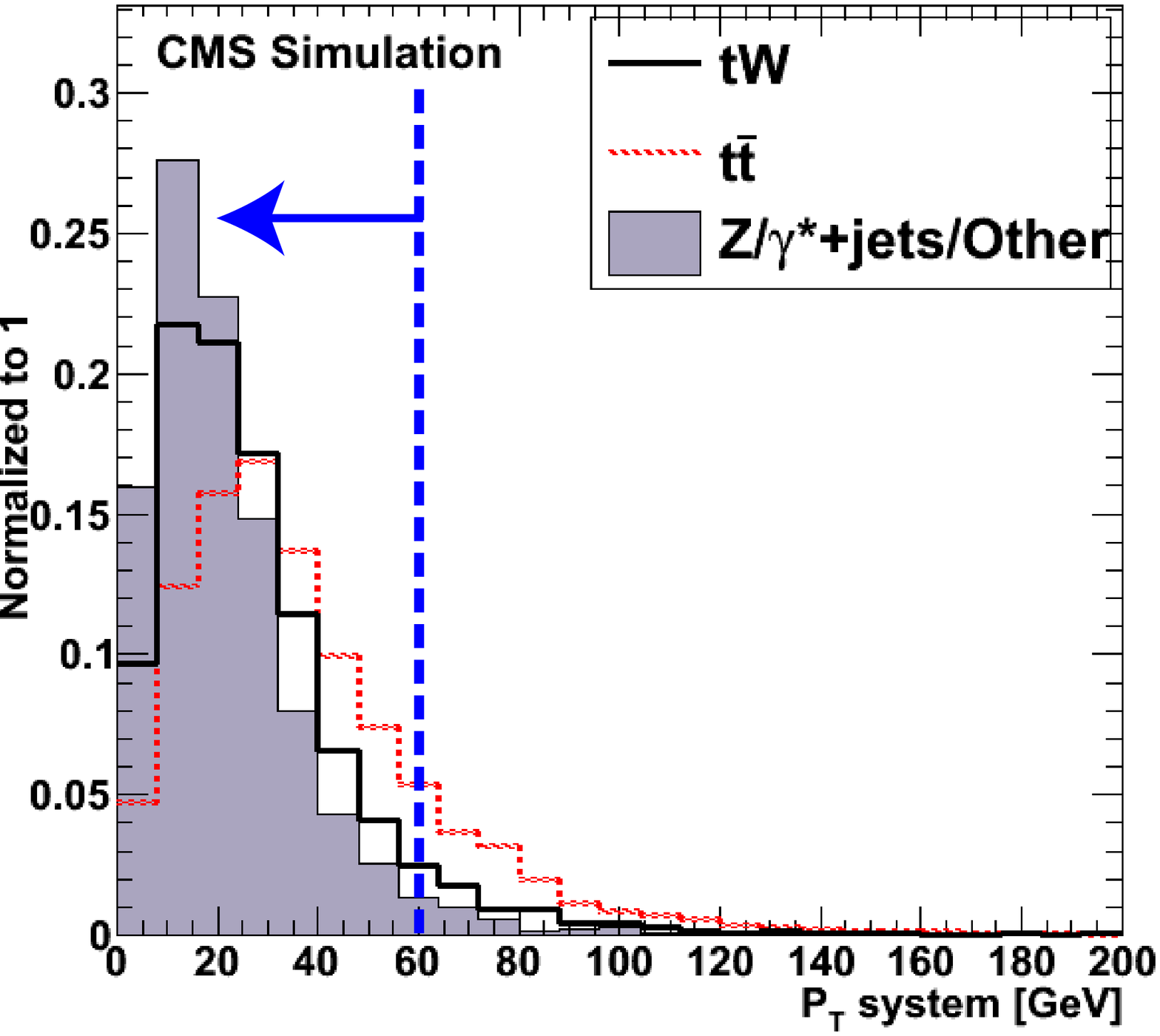}}
			\qquad
			\resizebox{0.45\columnwidth}{!}{
				\includegraphics{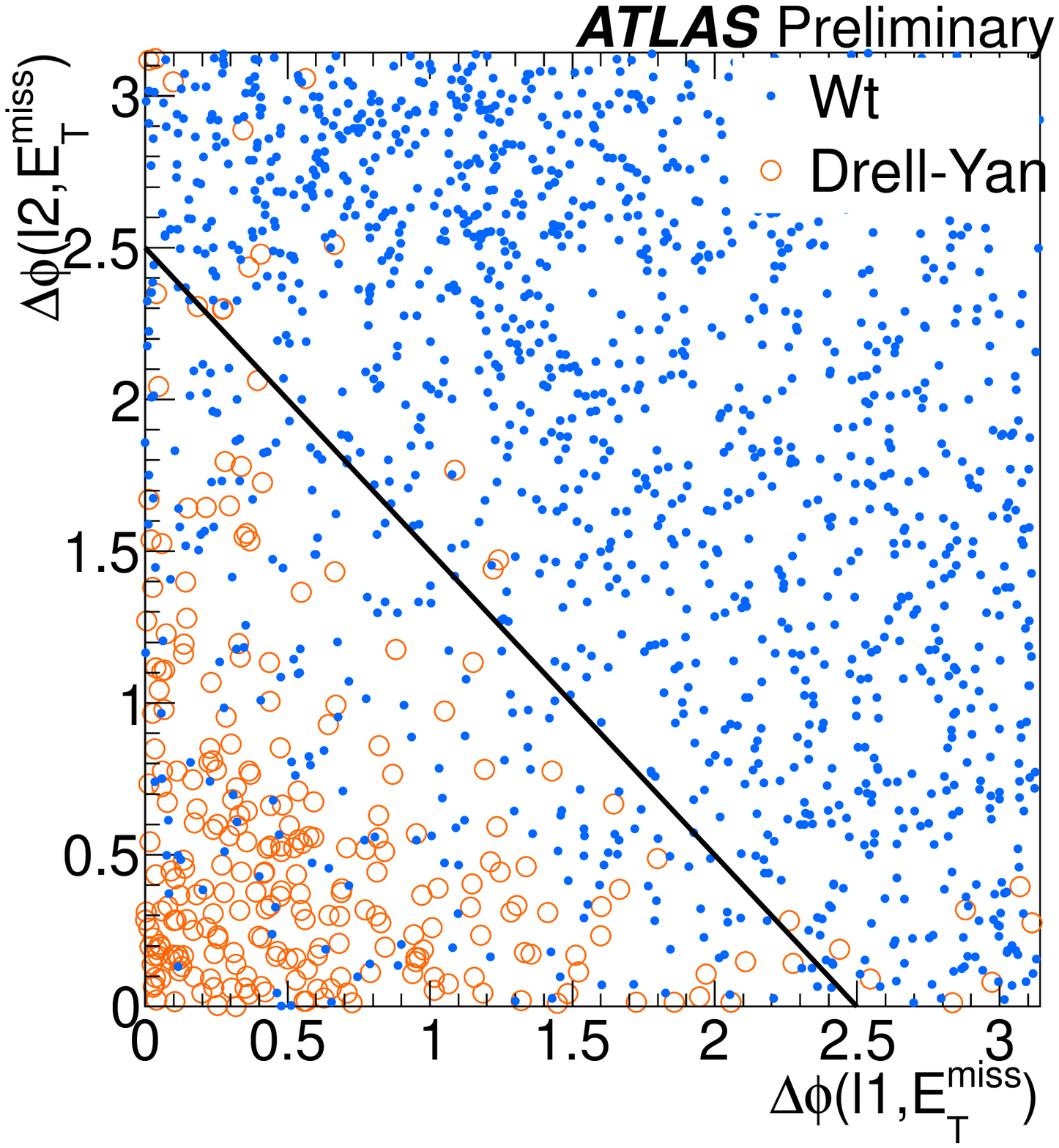}}
		\vspace{-0.5cm}
		\caption{Cuts to suppress \ttbar and Drell-Yan background in the ATLAS \protect\cite{ATL_Wt} and CMS  \protect\cite{CMS_Wt} Wt-channel analyses. \emph{Left:} Transverse momentum of the (jl\MET)-system $P_T^{system}$. \emph{Right:} Triangle cut in the $\Delta\phi(l_{1,2},\MET)$ plane.}
		\label{fig:Wt_cuts} 
	\end{center}
\end{figure}

\begin{figure}
	\begin{center}
		\resizebox{0.49\columnwidth}{!}{
			\includegraphics{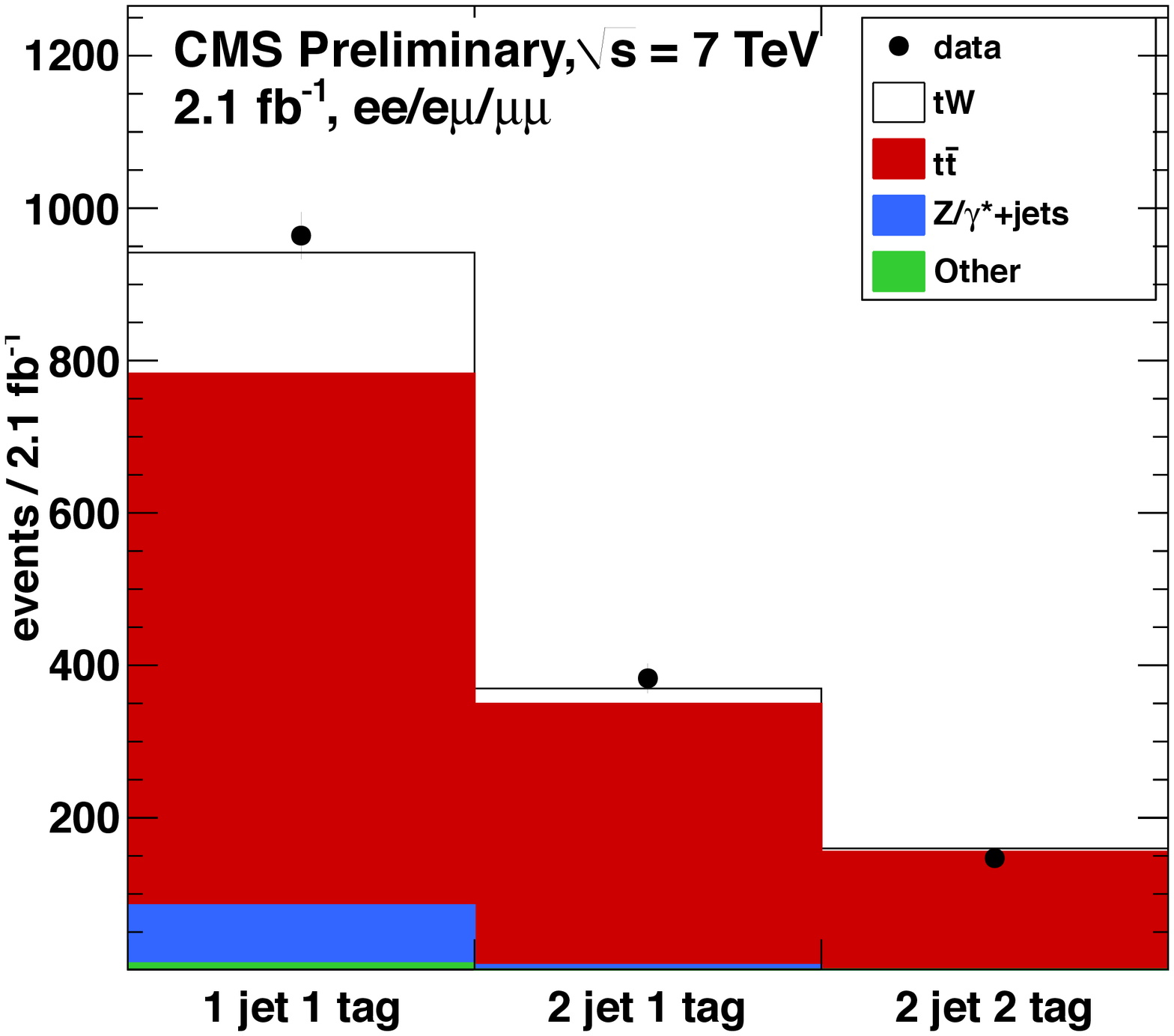}}
		\resizebox{0.49\columnwidth}{!}{
			\includegraphics{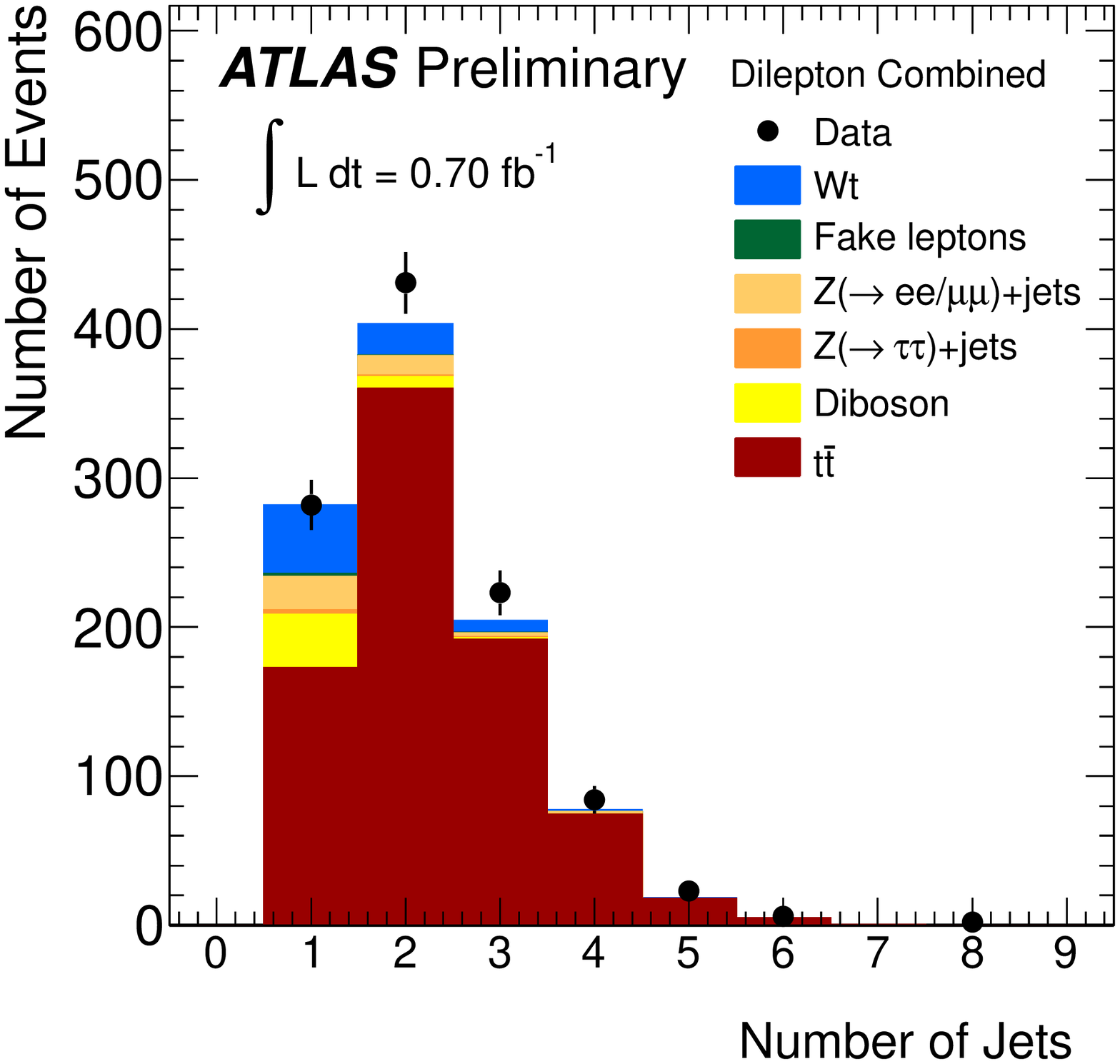}}
		\vspace{-0.5cm}
		\caption{Results of the Wt-channel analyses for ee, $\mu\mu$ and e$\mu$ channels combined. \emph{Left:} CMS \protect\cite{CMS_Wt} fits the three jet bins indicated simultaneously to constrain signal and backgrounds. \emph{Right:} ATLAS \protect\cite{ATL_Wt} uses the first bin as signal bin, the \ttbar background is estimated from the $\ge2$-jets sideband.}
		\label{fig:Wt_results} 
	\end{center}
\end{figure}

In the single top production channel with an associated \W boson, CMS has performed an analysis on 2.1~fb$^{-1}$ \cite{CMS_Wt}. ATLAS has analyzed the full 2010 dataset of 35~pb$^{-1}$ \cite{ATL_t_1} and updated this analysis with 700~pb$^{-1}$ of 2011 data \cite{ATL_Wt}. Both collaborations select the dilepton signatures only (ee, $\mu\mu$, e$\mu$) where both the associated \W and the \W stemming from the top decay leptonically. Due to that, exactly two opposite-sign leptons are required. ATLAS cuts on \mbox{\MET$>50$} GeV in all three signatures. CMS cuts on \mbox{\MET$>30$}~GeV in the ee and $\mu\mu$ signatures. In the e$\mu$ case, no cut on \MET is performed, but the total transverse mass in the event $H_T$ is required to be larger than 160~GeV. CMS asks for exactly one jet, which has to be identified as a \qb-jet, whereas ATLAS does not have a \qb-identification requirement on its one jet. To reduce the amount of background in the samples, events in the \Z boson mass window $81<M_{ee,\mu\mu}<101$~GeV are rejected in both analyses. To get rid of Drell-Yan background, CMS rejects events where the dilepton mass m$_{ll}$ is smaller than 20~GeV. ATLAS performs a triangle cut on the angle between the lepton and \MET in the transverse plane (see Fig.~\ref{fig:Wt_cuts}, \emph{right}). In order to reduce \ttbar background, CMS rejects events with an additional \qb-jet above 20~GeV and requires the transverse momentum of the system consisting of the two leptons, jet and \MET to be smaller than 60~GeV (cf. Fig.~\ref{fig:Wt_cuts}, \emph{left}). %???VERIFY!!!

Both ATLAS and CMS perform data-driven background estimates. The Drell-Yan background is estimated by CMS from the \Z-veto rejection region. ATLAS estimates this background with the ABCDEF method in the ($M_{ll}$, \MET) plane. The \ttbar background is estimated from the sideband, namely the $\ge\!2$-jets region. In contrast, CMS performs a simultaneous fit of the signal and two control regions, the 2-jet 1-tag and the 2-jet 2-tags samples. Results for both analyses are shown in Fig.~\ref{fig:Wt_results}.

The CMS analysis of 2.1~fb$^{-1}$ measures a cross section in the Wt-channel of $22^{+9}_{-7}$~pb where statistical an systematic uncertainties have been combined. This corresponds to a 2.7\,$\sigma$ deviation from the background-only hypothesis, where 1.8$\pm$0.9\,$\sigma$ are expected. The dominant systematic uncertainty comes from the jet energy scale and jet resolution. Also the factorization / normalization scale, initial and final state radiation and \qb-identification contribute significantly. The ATLAS analysis on the smaller dataset of 0.7~fb$^{-1}$ yields a cross section of $14\pm5^{+10}_{-9}$~pb. As this is consistent with a background-only hypothesis on the 
1.2\,$\sigma$ level, an upper bound on the Wt-channel cross section is derived. It is measured to be 39~pb, where 41~pb are expected. Dominant systematic uncertainty also in this case is the jet reconstruction including energy scale, resolution and efficiency. The generator choice is important as well. 
%Table~\ref{tab:Wt_syst} gives an overview of statistical and systematic uncertainties for this analysis. It can be seen, that
Although the analyzed dataset is still small compared to what is already available and will be provided in the future, the uncertainty is already dominated by systematics.

%%%%%%%%%%%%%%%%%%%%%%%%%%%%%%%%%%%%%%%%%%%%

\section{Measurement in the s-channel}
\label{sec:schannel}

\begin{figure}
	\begin{center}
	\resizebox{0.48\columnwidth}{!}{
		\includegraphics{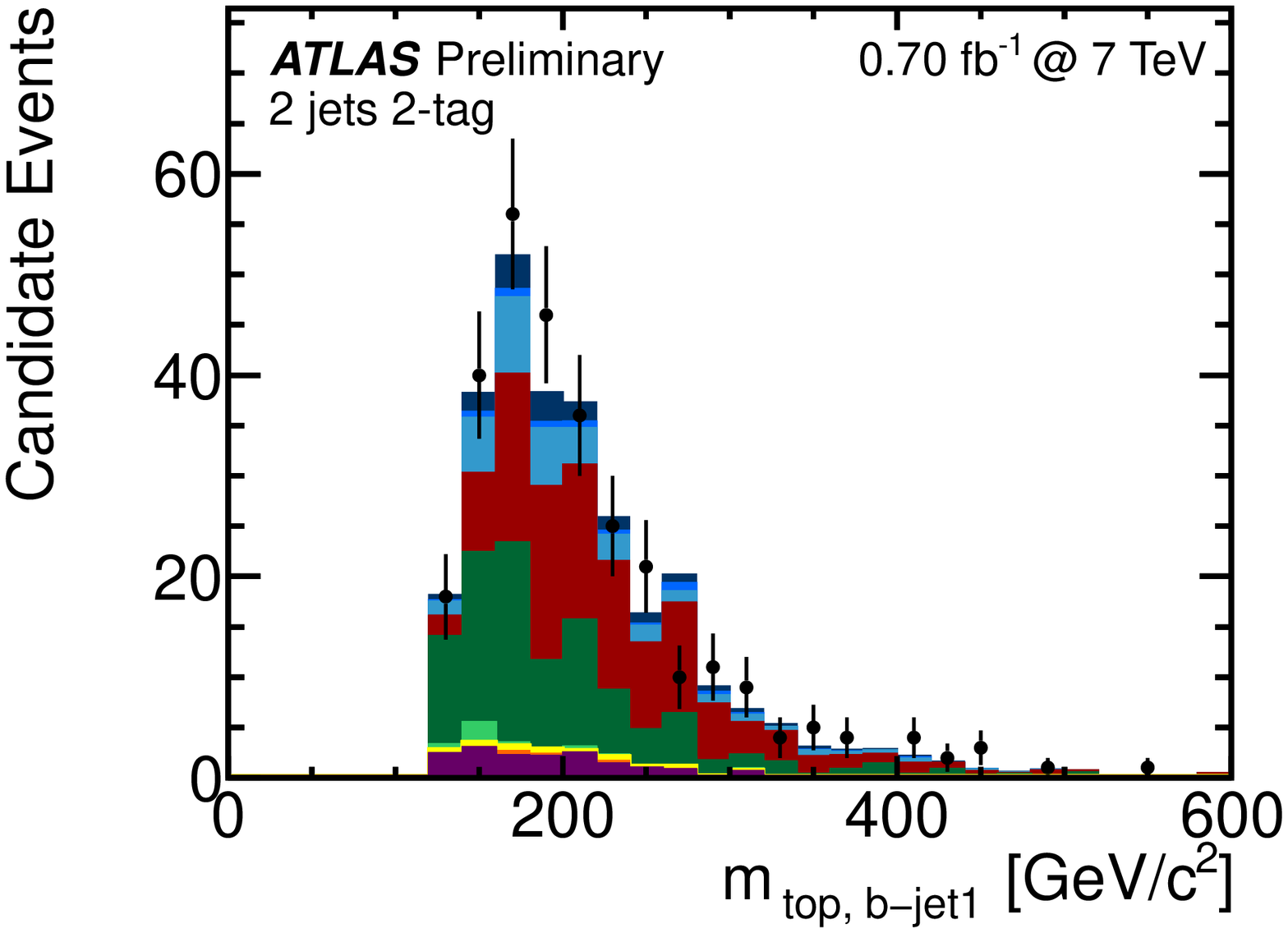}
	}
	\resizebox{0.51\columnwidth}{!}{
		\includegraphics{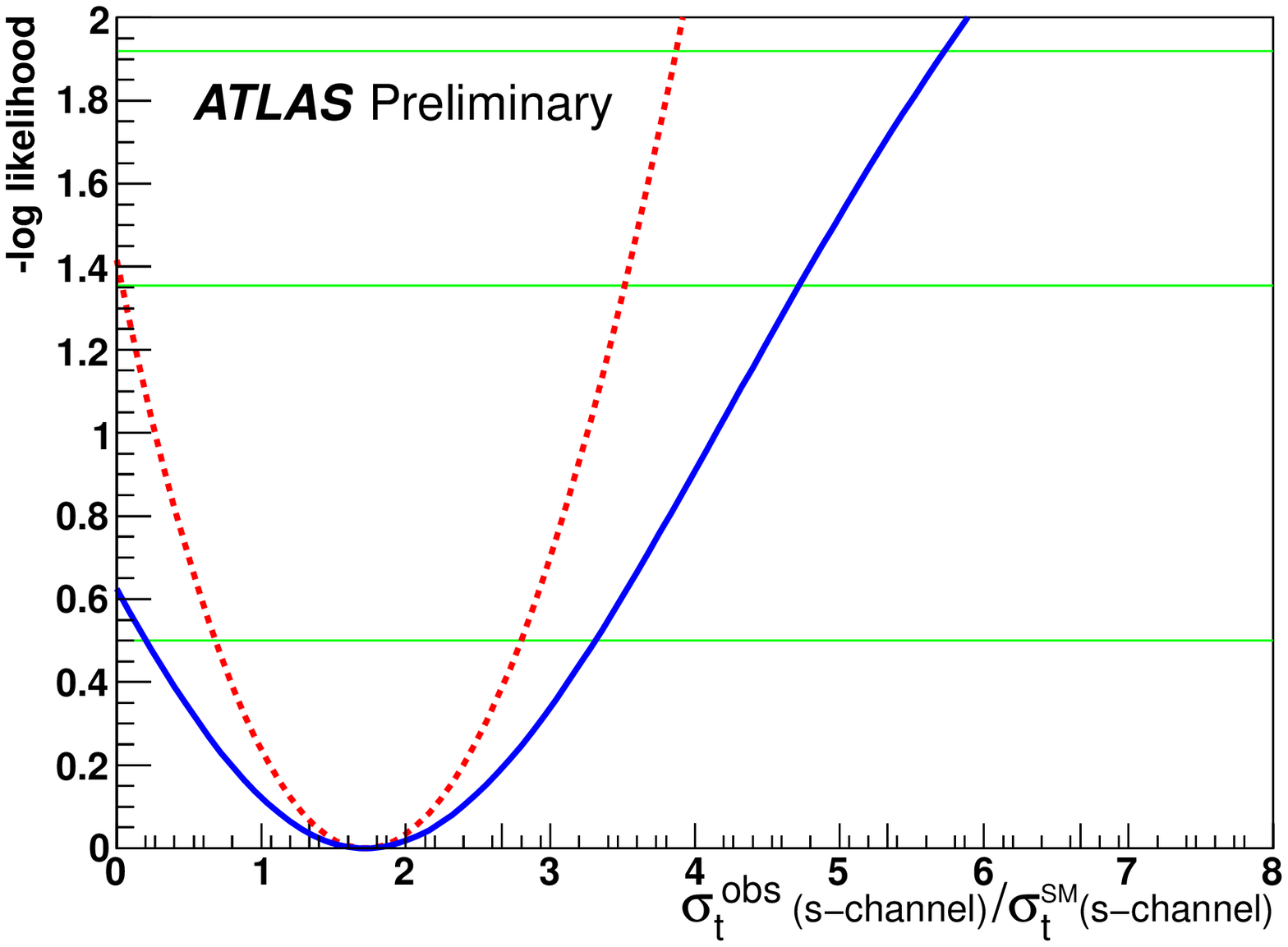}
	}
	\vspace{-0.7cm}
	\caption{Results of the ATLAS s-channel analysis \protect\cite{ATL_s}. \emph{Left:} Top mass distribution built with the leading \qb-jet. \emph{Right:} Negative log-likelihood distribution of the fit vs. ratio of observed over expected cross section. \emph{Red:} stat. uncertainty only, \emph{blue:} systematic uncertainty included.}
	\label{fig:ATL_s} 
	\end{center}
\end{figure}

The s-channel, which has the smallest cross section of the three single top production channels has been addressed only by ATLAS so far. The analysis is based on 0.7~fb$^{-1}$ \cite{ATL_s} corresponding to the same 2011 dataset as the t-channel and Wt-channel analyses. Again, only leptonic \W decays to electrons or muons are taken into account. the transverse momentum of the lepton and jets are required to be larger than 25~GeV as well as the \MET. The same cut on the transverse mass of the \W as in the t-channel analysis is performed ($\MTW>60GeV-\MET$). The QCD multijet background is estimated from data from the \MET distribution in a loose isolation lepton sample. The \Wjets background shape is taken from \MC simulation whereas the overall normalization and contributions of the different flavour subsamples is derived from data. Background from other top production channels, i.e. \ttbar and t-channel single top, is estimated from theory predictions. In a final selection step, exactly two jets identified as \b-jets are required. This brings $S\!/\!\sqrt{B}$ from 0.26 to 0.88. In addition, cuts on the mass of the reconstructed top quark and the first, respectively second, jet, the transverse momentum of the two jets combined, the transverse mass of the reconstructed \W and the distance in $\Delta R$ between the first jet and the lepton as well as the two jets are performed. A detailed list of these cuts is given in \cite{ATL_s}. This increases $S\!/\!\sqrt{B}$ to 0.98. A total of 296 events are selected, where 285$\pm$17 are predicted. Figure~\ref{fig:ATL_s}, \emph{left} shows the reconstructed top mass distribution build from l, \MET and the leading \qb-jet.

A likelihood fit (cf. Fig.~~\ref{fig:ATL_s}, \emph{right}) is performed that yields an upper limit on s-channel single top production of 20.5~pb, where uncertainties have been taken into account. The expected limit is 26.5~pb. The dominant systematic uncertainties are roughly equal in size from \MC generator choice, luminosity estimate, QCD background normalization and \qb-identification. This analysis is still statistically limited however the systematic uncertainty already contributes significantly.

 %LEGEND!!!

%%%%%%%%%%%%%%%%%%%%%%%%%%%%%%%%%%%%%%%%%%%%

\section{Search for Flavour Changing Neutral Currents}
\label{sec:FCNC}

Flavour Changing Neutral Currents (FCNC) in single top production are highly suppressed in the Standard Model. However, many models beyond the Standard Model (BSM) predict enhanced branching ratios for this process, which can be seen in Fig.~\ref{fig:feyn_FCNC}. Still, the expected signal is very small compared to background processes and only multivariate techniques seem to be promising in separating them. 

\begin{figure}
	\begin{center}
	\resizebox{0.45\columnwidth}{!}{
		\includegraphics{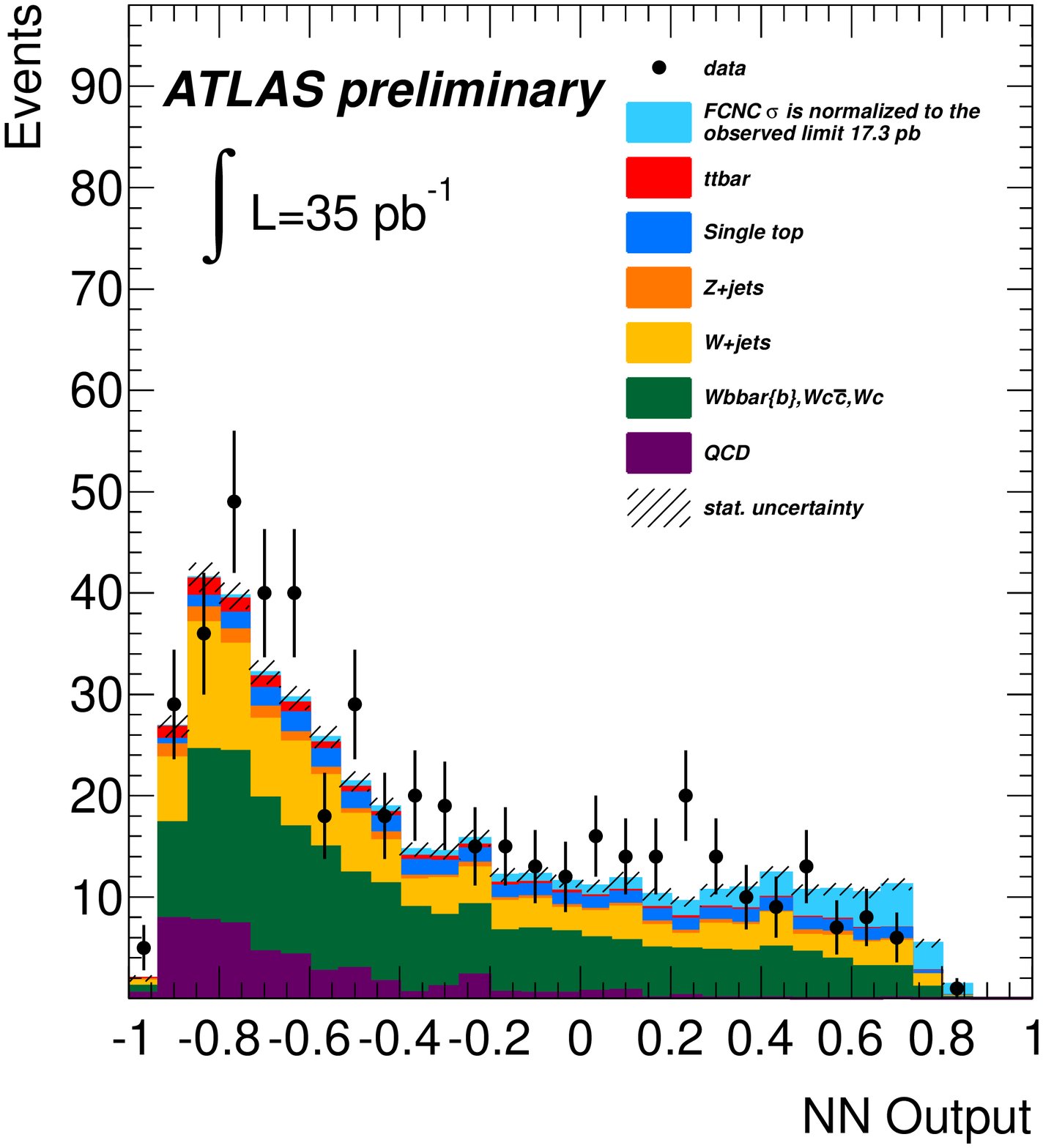}
	}
	\qquad
	\resizebox{0.45\columnwidth}{!}{
		\includegraphics{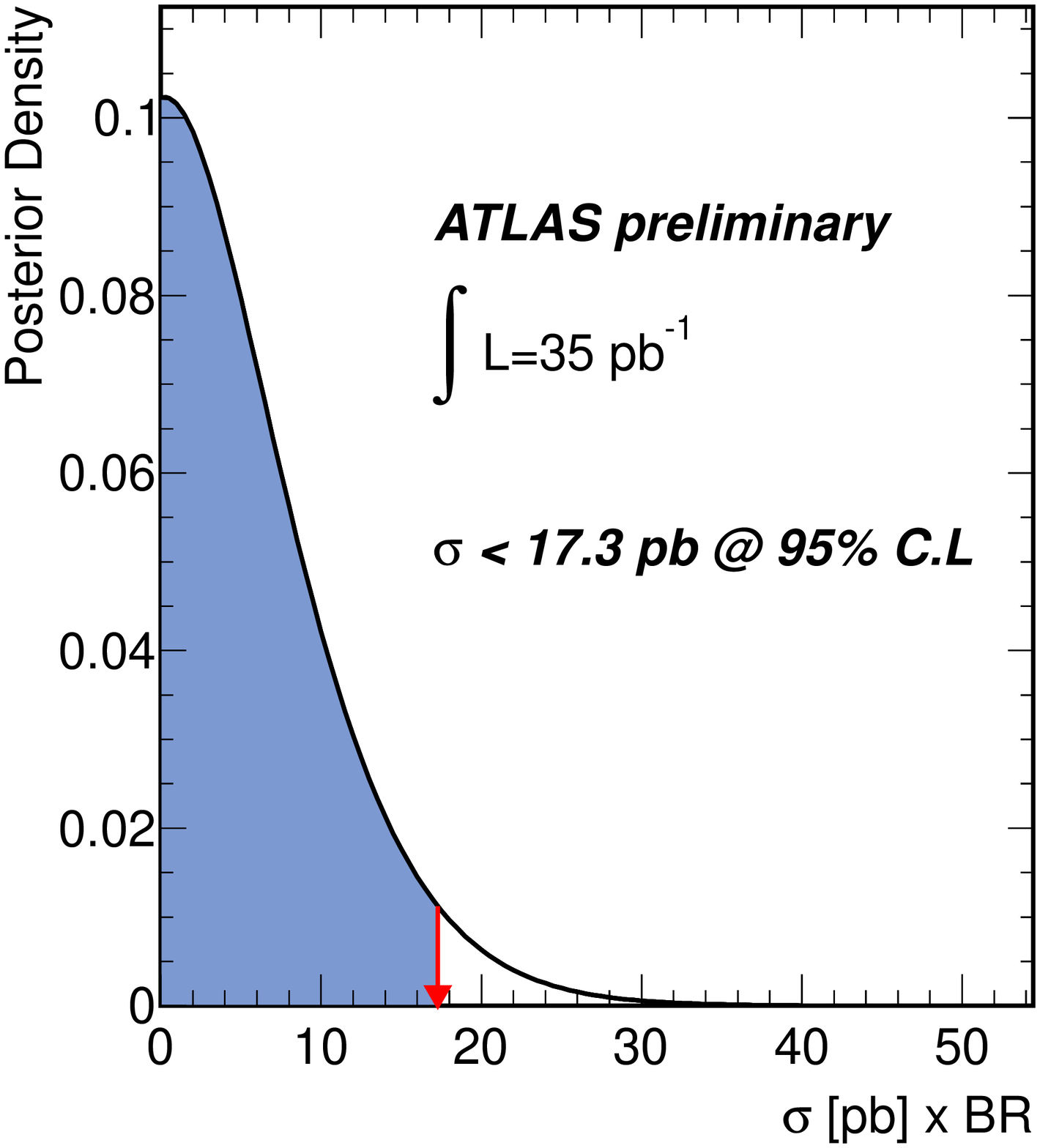}
	}
	\vspace{-0.7cm}
	\caption{Results of the ATLAS search for FCNC \protect\cite{ATL_FCNC}. \emph{Left:} NN output distribution including a signal sample scaled to the cross section of the observed upper limit. \emph{Right:} Posterior probability function of the observed events including all systematic uncertainties. The red arrow marks the value measured in data.}
	\label{fig:ATL_FCNC} 
	\end{center}
\end{figure}

ATLAS has performed an analysis on 35~pb$^{-1}$ \cite{ATL_FCNC}, the full 2010 dataset, which is the first search for BSM processes in single top signatures. As in the previously described cross section analyses, only leptonic \W boson decays are considered, with an electron or muon above 25~GeV in the final state. Cuts on the missing transverse energy and the transverse mass of the \W require $\MET>35$~GeV (20~GeV) and $\MTW>25$~GeV (60~GeV$-\MET$) in the electron (muon) channel. Events with exactly one jet, which has to be identified as a \qb-jet, with a transverse momentum above 25~GeV are selected. A Neural Network is used to separate the signal from \ttbar, SM single top production, \mbox{\W/\Z+jets} and QCD multijet background. The result is shown in Fig.~\ref{fig:ATL_FCNC}, \emph{left}. No evidence for a signal has been found and an upper limit on a possible FCNC contribution has been derived (cf. Fig~\ref{fig:ATL_FCNC}, \emph{right}). On a 95\% confidence level (C.L.), the cross section is smaller than 17.3~pb. This is in good agreement with the expected limit of 17.4~pb.

%%%%%%%%%%%%%%%%%%%%%%%%%%%%%%%%%%%%%%%%%%%%

\section{Conclusions}
\label{sec:conclusions}

Analyses of single top production have been performed by the ATLAS and CMS collaborations in the t-channel and Wt-channel on 2010 and 2011 data. The t-channel has been established by ATLAS with a significance of 7.6~$\sigma$ in a data sample of 700~pb$^{-1}$. The cross section was measured as $90\pm9^{+31}_{-20}$~pb, with $64\pm3$~pb being expected in the Standard Model. This is the first observation of single top production at the LHC. CMS has analyzed the full 2010 dataset of 36~pb$^{-1}$ and found evidence for t-channel single top production on the 3.5~$\sigma$ level. 

The Wt-channel has been measured for the first time as this channel was not accessible at the Tevatron. CMS found a cross section of $22^{+9}_{-7}$~pb with $16\pm1$~pb expected. Although this analysis has been performed on the largest dataset of 2.1~fb$^{-1}$ in all single top analyses, the result is still compatible with the background-only hypothesis on the 2.7~$\sigma$ level. ATLAS, on a smaller dataset of 0.7~fb$^{-1}$, found a deviation of only 1.2~$\sigma$. However, it should be noted that a dataset of 5~fb$^{-1}$ is already available in both collaborations, so that evidence in this channel can be expected in the near future. Both collaborations have shown that with the larger statistics available at the LHC simple cut-based analyses are possible in both channels and multivariate techniques, which were used at the Tevatron, are not a necessary prerequisite anymore.

The s-channel, with the smallest expected Standard Model cross section of $4.6\pm0.2$~pb, has been analyzed by ATLAS and a cross section limit of 26.5~pb at 95\% C.L. was derived. Here, much more statistics are needed before evidence can be claimed.

A search for FCNC in single top production was performed by ATLAS for the first time and a cross section limit of 17.3~pb at 95\% C.L. could be set. This analysis was performed on only 35~pb$^{-1}$ of data, so that a reduced limit based on the larger 2011 dataset can be expected in the next round of analyses. This result shows that searches in single top signatures are already possible with the available datasets and the search program should be extended to more new phenomena signatures in the future.

%%%%%%%%%%%%%%%%%%%%%%%%%%%%%%%%%%%%%%%%%%%%
%


\begin{thebibliography}{}

\bibitem{xsec_1}
N. Kidonakis, Phys.Rev. \textbf{D81}, (2010) 054028
\bibitem{xsec_2}
N. Kidonakis, Phys.Rev. \textbf{D82}, (2010) 054018
\bibitem{xsec_3}
N. Kidonakis, Phys.Rev. \textbf{D83}, (2011) 091503

\bibitem{CMS_t}
The CMS Collaboration, PRL \textbf{107}, (2011) 091802
\bibitem{ATL_t_1}
The ATLAS Collaboration, ATLAS-CONF-2011-027, http://cdsweb.cern.ch/record/1336762
\bibitem{ATL_t_2}
The ATLAS Collaboration, ATLAS-CONF-2011-088, http://cdsweb.cern.ch/record/1356197
\bibitem{ATL_t_3}
The ATLAS Collaboration, ATLAS-CONF-2011-101, http://cdsweb.cern.ch/record/1369217

\bibitem{CMS_Wt}
The CMS Collaboration, CMS-PAS-TOP-11-022, http://cdsweb.cern.ch/record/1385552

\bibitem{ATL_Wt}
The ATLAS Collaboration, ATLAS-CONF-2011-104, http://cdsweb.cern.ch/record/1369829

\bibitem{ATL_s}
The ATLAS Collaboration, ATLAS-CONF-2011-118, http://cdsweb.cern.ch/record/1376410

\bibitem{ATL_FCNC}
The ATLAS Collaboration, ATLAS-CONF-2011-061, http://cdsweb.cern.ch/record/1345084
% etc

\end{thebibliography}
\end{document}